\documentclass[12pt]{article}

\usepackage{graphicx}
\usepackage{times}
\usepackage{amssymb}
\usepackage{fancyhdr}

\usepackage{epsfig} 
\usepackage{multirow} 

\begin{document}

\renewcommand{\multirowsetup}{\centering} 

\title{Prediction and predictability of global epidemics:\\
the role of the airline transportation network}

\author{Vittoria Colizza$^1$, Alain Barrat$^2$, Marc Barth{\'e}lemy$^1$, and Alessandro Vespignani$^1$}
\maketitle

\begin{center}
\small{$^1$ School of Informatics and Biocomplexity Center,
Indiana University, Bloomington, 47406, IN, USA\\ $^2$ UMR du
CNRS 8627,LPT B{\^a}timent 210, Universit{\'e} de Paris-Sud, 91405 ORSAY
Cedex - France}
\end{center}


\begin{abstract}
The systematic study of large-scale networks has unveiled the 
ubiquitous presence of connectivity patterns 
characterized by large scale heterogeneities and unbounded statistical 
fluctuations. These features affect dramatically the behavior of the 
diffusion processes occurring on networks, determining the ensuing 
statistical properties of their evolution pattern and dynamics.  
In this paper, 
we investigate the role of the large scale properties of the airline 
transportation network in determining the global evolution of 
emerging disease. 
We present a stochastic computational framework for the forecast 
of global epidemics that considers the complete 
world-wide air travel infrastructure complemented with census 
population data. We address two basic issues in global epidemic
modeling: i) We study the role of  the large scale properties of the airline
transportation network in determining the global diffusion pattern of
emerging diseases; ii) We evaluate the reliability of forecasts and
outbreak scenarios with respect to the intrinsic stochasticity
of disease transmission and traffic flows. In order to address these 
issues we define a set of novel quantitative measures able 
to characterize the level of heterogeneity and predictability of 
the epidemic pattern. These measures may be used for the analysis 
of containment policies and epidemic risk assessment.\\
\end{abstract}

The mathematical modeling of epidemics has often dealt with the
predictions and predictability of outbreaks in real populations with
complicated social and spatial structures and with heterogeneous
patterns in the contact
network~\cite{May:1992,Hethcote:1984,Morris:1996,Keeling:1999,Pastor:2001,May:2001,Meyers:2005,Ferguson:2003}.
All these factors have led to sophisticate modeling 
approaches including disease realism, meta-population grouping, 
stochasticity and more recently to agent based numerical simulations that 
recreate entire populations and their dynamics at the scale of the 
single individual \cite{Chowell:2003,Eubank:2004}.  
In many instances however the introduction of 
the inherent complex features and emerging
properties~\cite{Barabasi:2000,Doro:2003,Pastorbook:2003} of the
network in which epidemics occur
implies the breakdown of 
standard homogeneous approaches~\cite{Pastor:2001,May:2001}
 and calls for a systematic
investigation of the impact of the detailed system's characteristics
in the evolution of the epidemic outbreak. 
These considerations are particularly relevant in the study of the
geographical spread of epidemics where the various long-range
heterogeneous connections typical of modern transportation networks
naturally give rise to a very complicated evolution of epidemics characterized
by heterogeneous and seemingly erratic
outbreaks~\cite{Cohen:2000,Cliff:2004}, as recently documented in the SARS
case~\cite{SARS}.  
In this context, air-transportation represents a major channel of epidemic
propagation, as pointed out in the modeling approach to global
epidemic diffusion of Rvachev and Longini~\cite{Longini:1985} 
capitalizing on previous studies on the russian airline 
network~\cite{Baroyan:1969}.
Similar modeling approaches, even if limited by a very partial knowledge of
the world-wide transportation network, have been used to study
specific outbreaks such as pandemic
influenza~\cite{Longini:1988,Grais:2003,Grais:2004},
HIV~\cite{Flahault:1991}, and very recently SARS~\cite{Hufnagel:2004}.  The
availability of the complete world-wide airport network (\emph{WAN}) 
dataset and the recent extensive studies of its
topology~\cite{Barrat:2004,Amaral:2004} are finally allowing a full
scale computational study of global epidemics. In the following we 
will consider for the first time a global 
stochastic epidemic model including the full International Air 
Transport Association (IATA) \cite{IATA} database, aiming at a detailed 
study of the interplay among the network structure and the stochastic 
features of the infection dynamics in defining the global spreading 
of epidemics. In particular, while previous studies have generally been focused
in the {\em a-posteriori} analysis of real case studies of global epidemics,
the large scale modeling presented here allows us to address more
basic theoretical issues such as the statistical properties of
the epidemic pattern and the effect on it of the complex architecture 
of the underlying transportation network. Finally such a detailed level of description allows for the first
time the quantitative assessment of the reliability 
of the obtained forecast with respect to the stochastic nature of the
disease transmission and travel flows, the outbreak initial conditions and 
the network structure.\\ 
  
\noindent  
{\bf The air transportation network}
  
\begin{figure}
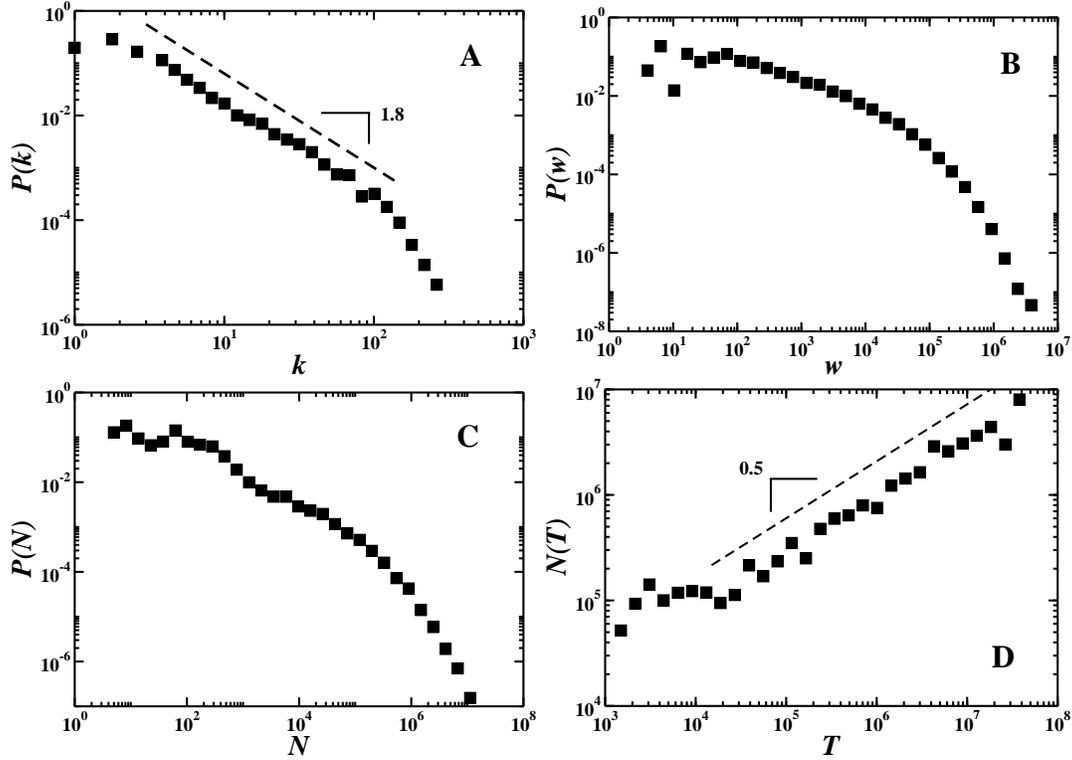
 

\centerline{ 
\includegraphics*[width=7cm]{Pk}   
\includegraphics*[width=7cm]{Pw}  
}

\centerline{ 
\includegraphics*[width=7cm]{PN}   
\includegraphics*[width=7cm]{NvsT} 
} 
\caption{Properties of the world-wide airport network. Statistical fluctuations
are observed over a broad range of length scales. {\bf (A)} The degree
distribution $P(k)$ follows a power-law behavior on almost two decades
with exponent $1.8\pm0.2$. {\bf (B)} The distribution of the weights
(fluxes) is skewed and heavy-tailed.  {\bf (C)} The distribution of
populations is heavy-tailed distributed, in agreement with the
commonly observed Zipf's
law~\cite{Zipf:1949}.  {\bf (D)} The city population varies with the
traffic of the corresponding airport as $N \sim T^\alpha$ with
$\alpha\simeq0.5$, in contrast with the linear behavior postulated in
previous works~\cite{Hufnagel:2004}.  } 
\label{fig:wan} 
\end{figure}

In the following we use the International Air Transport 
Association (IATA) \cite{IATA} database containing the world list of 
airport pairs connected by direct flights and the number of available 
seats on any given connection for the year 2002. The resulting 
world-wide air-transportation network (\emph{WAN}) is therefore a 
weighted graph comprising $V=3880$ vertices denoting airports and 
$E=18810$ weighted edges whose weight $w_{j\ell}$ accounts for the 
passenger flow between the airports $j$ and $\ell$. This dataset has 
been complemented by the population $N_j$ of the large metropolitan 
area served by the airport as obtained by different sources. The final
network dataset contains the 3100 largest airports, 17182 edges
(accounting for 99\% of the worldwide traffic)  and the respective
urban population data. 
The obtained network is highly heterogeneous both in the connectivity 
pattern and the traffic capacities (see Fig.~\ref{fig:wan}).  
  The probability distributions 
that an airport $j$ has $k_j$ connections (degree) to other airports and 
handles a number $T_j=\sum_{\ell}w_{j\ell}$ of 
passengers (traffic) exhibit heavy-tails and 
very large statistical fluctuations~\cite{Amaral:2004,Barrat:2004}. 
Analogously, the probability that a connection has a traffic $w$ is 
skewed and heavy-tailed. Finally, we associate to each airport a city 
whose population $N$ is heavy-tailed distributed in agreement with the 
general result of Zipf's law for the city size~\cite{Zipf:1949}.  More 
strikingly, these quantities appear to have non-linear associations 
among them.  This is clearly shown by the behavior relating the 
traffic handled by each airport $T$ with the corresponding number of 
connections $k$ that follows the non-linear form $T\sim k^\beta$ with 
$\beta\simeq1.5$ \cite{Barrat:2004}.  Analogously, the city population 
and the traffic handled by the corresponding airport follows the 
non-linear relation $N\sim T^\alpha$ with $\alpha\simeq0.5$ in 
contrast with the linear behavior assumed in previous 
analysis~\cite{Hufnagel:2004}. The presence of broad statistical 
distributions and non-linear relations among the various quantities 
indicate a possible major impact in the ensuing disease spreading 
pattern.\\
\\

\noindent  
{\bf Modeling global epidemics}  
 
As a basic element of our modeling approach we
assume the basic standard compartmentalization 
in which each individual can only 
exist in one of the discrete states such as susceptible (S), 
latent (L), infected (I),  permanently recovered (R), etc. 
In each city $j$ the population is
$N_j$ and $X^{[m]}_j(t)$ is the number of individuals in the class
$[m]$ at time $t$. By definition it follows that $N_j=\sum_m
X^{[m]}_j(t)$. In each city $j$ the individuals
 are allowed to travel from one city to another by 
means of the airline transportation network
and to change compartment because of the infection dynamics in each city,
similarly to the models in refs.~\cite{Longini:1985,Grais:2003,Flahault:1991}
 and 
the stochastic generalization of ref.~\cite{Hufnagel:2004}.\\ 
\\

{\it Transport operator}. The dynamics of individuals due to travels 
between cities is described by the transport operator 
$\Omega_j(\{X^{[m]}\})$ representing the net balance of individuals in a 
given class $X^{[m]}$  that entered and left each city 
$j$. This operator is a function of the traffic 
flows $w_{j \ell}$ per unit time, the city populations $N_j$, 
and might also include transit passengers 
on connecting flights.  
The number of passengers of each category 
traveling from a city $j$ to a city $\ell$ is an integer random 
variable, in that each of the $X^{[m]}_j$ potential travellers has a 
probability $p_{j\ell}=w_{j\ell}\Delta t/N_j$ to go 
from $j$ to $\ell$ in the time interval $\Delta t$.  In 
each city $j$ the numbers of passengers $\xi_{j\ell}$ traveling on each 
connection $j\to\ell$ at time $t$ define a set of stochastic variables 
which follow the multinomial distribution 
\begin{eqnarray}  
\nonumber 
P(\{\xi_{j\ell}\})=\frac{X^{[m]}_j!}{\left( 
X^{[m]}_j-\sum_\ell\xi_{j\ell}\right)!  
\prod_\ell\xi_{j\ell}!}    
\prod_\ell p_{j\ell}^{\xi_{j\ell}} 
\times \\ 
\times 
\left(1-\sum_\ell p_{j\ell}\right)^{(X^{[m]}_j-\sum_\ell\xi_{j\ell})} 
, 
\label{multi} 
\end{eqnarray}  
where $\left(X^{[m]}_j-\sum_\ell\xi_{j\ell}\right)$ identifies the number of 
non traveling individuals, and we use standard numerical  
subroutines to generate random numbers  
of travellers following these distributions. 
 
The transport operator in each city $j$ is therefore written as 
\begin{equation}   
\Omega_j(\{X^{[m]}\})=\sum_\ell(\xi_{\ell j}(X^{[m]}_\ell) - \xi_{j\ell}(X^{[m]}_j)), 
\label{tran_stoc}  
\end{equation}   
where the mean and variance of the stochastic variables are 
$\langle\, \xi_{j\ell}(X^{[m]}_j)\,\rangle\,=\,p_{j\ell}X^{[m]}_j$ and
$\textrm{Var}(\xi_{j\ell}(X^{[m]}_j))=p_{j\ell}(1-p_{j\ell})X^{[m]}_j$. 
In addition, since the traffic flows are expressed as the number of
available seats on a given connection, we have to consider that the 
transport operator is in  
general affected by fluctuations due to an occupancy rate of the  
airplanes not equal to $1$. This introduces a further source of noise since   
we have to consider that on each connection $(j,\ell)$ the flux of passengers  
at each time $t$ is given by a stochastic variable  
\begin{equation}   
\tilde{w}_{j\ell}=w_{j\ell}[\alpha+\eta(1-\alpha)]  
\end{equation}   
where $\alpha=0.7$ corresponds to the average occupancy rate of 70\%
provided by official statistics \cite{IATA} and $\eta$ is a random 
number drawn uniformly in the interval $[-1,1]$ at each time step.\\ 
  
{\it Infection dynamics}. 
The dynamics of the individuals $X^{[m]}$ between the different
compartments depends on the specific disease considered.
In compartmental models there are two possible elementary processes
ruling the disease dynamics. The first class of process refers to the
spontaneous transition of one individual from one compartment $[m]$
to another compartment $[h]$.
Processes of this kind are the spontaneous recovery of infected
individuals ($I\to R$) or the passage from a latent condition to an
infectious one ($L\to I$) after the incubation period.  In this case
the variation in the number of individuals $X^{[m]}$ is simply given
by $\sum_h \nu_h^m a_h X^{[h]}$, where $a_h$ is the rate of transition
from the class $[h]$ and $\nu_h^m \in \{-1,0,1\}$ is the change in the
number of $X^{[m]}$ due to the spontaneous process from or to the
compartment $[h]$. The second class of processes refers to binary 
interaction among
individuals such as the contagion of one susceptible in interaction
with an infectious ($S+I\to 2I$).  In the homogeneous assumption, 
the rate of variation of individuals $X^{[m]}$ is given by $\sum_{h,g} \nu_{h,g}^m
a_{h,g}N^{-1} X^{[h]}X^{[g]}$, where $a_{h,g}$ is the rate of
transition rate of the process and $\nu_{h,g}^m \in\{ -1,0,1\}$ the
change in the number of $X^{[m]}$ due to the interaction. The factor
$N^{-1}$, where $N$ is the number of individuals, stems from the fact
that the above expression considers the homogeneous approximation in
which the probability for each individual of class $[h]$ to interact
with an individual of class $[g]$ is simply proportional to the
density $X^{[g]}/N$ of such individuals (note that it is however possible 
to consider other cases~\cite{May:1992}).\\

{\it Stochastic formulation of the global spreading model}. 
In order to go beyond the usual deterministic approximations, 
in each city we work directly with
the master equations for the processes described above~\cite{Hufnagel:2004}
 and under the
assumption of large populations we obtain the Langevin equations in
which we associate to each reaction process a noise term with
amplitude proportional to the square root of the reaction
term~\cite{Gardiner,Gillespie,MarroDickman}.  The
epidemic Langevin equations are coupled among them by the stochastic transport
operator that describes movements of individuals from one city to 
another and can be
numerically solved by considering the discretized evolution equations~\cite{Gardiner,Gillespie,MarroDickman} 
for small time steps $\Delta t$ 
that read
\begin{eqnarray}
X^{[m]}_j(t+\Delta t)-X^{[m]}_j(t)=\sum_{h,g}N_{j}^{-1} \nu_{h,g}^m
a_{h,g} X_j^{[h]}(t) X_j^{[g]}(t) \Delta t+ & & {}
\nonumber\\
   {}+ \sum_h \nu_h^m
a_h X_j^{[h]}(t)\Delta t + \sum_{h,g} \nu_{h,g}^m
\sqrt{a_{h,g} N_{j}^{-1} X_j^{[h]} X_j^{[g]}\Delta t}\ \eta_{h,g} + & & {}
\nonumber\\
 {}+ \sum_h \nu_h^m
\sqrt{a_h X_j^{[h]}\Delta t}\ \eta_h +\Omega_j(\{X^{[m]}\}), & &
\label{eq:discret}
\end{eqnarray} 
where $\eta_{h,g}$ and $\eta_{h}$ are statistically independent
Gaussian random variables with zero mean  and unit variance  and
$\Omega_j(\{X\})$ is the stochastic travel operator (defined in
the previous paragraph) depending on the traveling probabilities (obtained
from the IATA dataset) $p_{j\ell}=w_{j\ell}\Delta t/N_j$. 
The model is thus a compartmental system of differential  
equations which can be numerically integrated. 
It is worth mentioning, however, that 
the standard integration of these equations by using Cauchy-Euler 
methods leads to a well-known technical problem and specific techniques 
must be used to avoid an asymmetric truncation of the noise 
terms~\cite{Dickman:1994}.\\

\noindent
{\bf The SIR dynamics}. 

Global epidemic forecast would be extremely relevant in the case of the 
emergence of a new pandemic influenza that in general spreads 
rapidly with substantial transmission occurring before the onset 
of case-defining symptoms. In the following we adopt a minimal 
model for a pandemic spread  to provide a 
general discussion that is not hindered by the use of very complicate 
disease transmission mechanisms. Specific characteristics such as 
latency, incubation and seasonal effects of the disease can be however
easily implemented in the present framework~\cite{Longini:1985}. 
The analysis of two case studies and the comparison of the 
forecasts obtained with the present approach  and the real data 
will be presented elsewhere. Here, we use the very simplistic 
approximation of the susceptible-infected-removed (SIR) dynamics 
in which a fully mixed population is assumed within each city.
In each city $j$ the population 
$N_j$ is given by $N_j=S_j(t)+I_j(t)+R_j(t)$, where $S_j(t)$, $I_j(t)$ 
and $R_j(t)$ represent the number of  susceptible, 
infected and recovered individuals at time $t$, respectively.
The epidemic evolution is 
governed by the basic dynamical evolution of the SIR model where the 
probability that a susceptible individual acquires the infection from 
any given infected individual in the time interval $dt$ is 
proportional to $\beta dt$, where $\beta$ is the transmission 
parameter that captures the aetiology of the infection process.  At 
the same time, infected individuals recover with a probability $\mu 
dt$, where $\mu^{-1}$ is the average duration of the infection. By
considering the three compartments $S, I$ and $R$ in
Eq.(\ref{eq:discret}) and plugging in $a_{I,S}=\beta$,  
$a_{I}=a_R=\mu$ and the corresponding parameters $\nu_{I,S}^I=1$ 
and $\nu_I^R=-\nu_I^I=1$ it is possible to obtain explicitly 
 the $3100\times 3$
differential equations whose integration provides the disease
evolution in every urban areas corresponding to an airport. 
Results shown in the following are obtained with the global spreading model
based on an SIR dynamics within each city.
 
\begin{figure*} 
\centerline{ 
\includegraphics*[width=0.9\textwidth,angle=90]{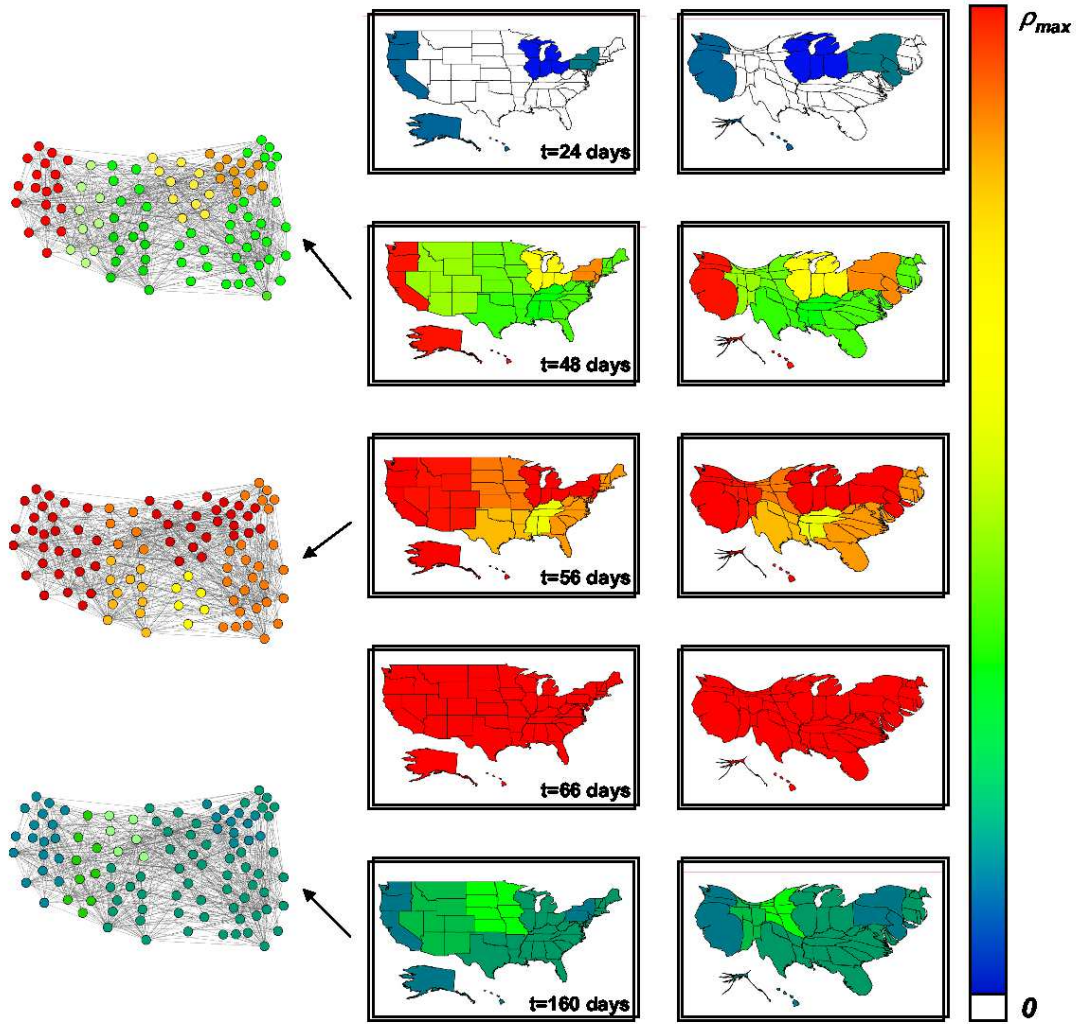} 
} 
\vspace{-.5cm}
\caption{Geographical  
representation of the disease evolution in the US for an epidemics
starting in Hong Kong based on an SIR dynamics within each city. 
States are collected according to the nine
influenza surveillance regions. The color code corresponds to the
prevalence in each region, from 0 to the maximum value reached ($\rho_{max}$).
On the left the original US maps are shown, while on the right
we provide the corresponding cartograms obtained by rescaling each
region according to its population. Three representations of the airport 
network restricted to the US  are also shown, in correspondance to 
 three different snapshots.
 The nodes represent the 100 airports in the US with highest 
traffic $T$; the color is assigned in accordance to the color code adopted 
for the maps.  
} 
\label{fig:maps} 
\end{figure*}

At first instance, it is possible to
monitor standard epidemiological quantities such as the level of infected  
individuals, the morbidity and the prevalence at different granularity  
levels; i.e. country, state or administrative regions.
In Fig.~\ref{fig:maps} we show the dynamical evolution in the US of 
an epidemic starting in Hong-Kong. The evolution of epidemic outbreaks 
is monitored by recording at each time step (1 day) the density of 
individuals in each class ($S$, $I$, $R$) present in each city. The 
parameters $\beta$ and $\mu$ are chosen according to~\cite{Grais:2004}
in order to use biologically sound  values
and kept constant during the 
evolution (different values do not lead to different overall 
conclusions). This amounts to 
assume that no restrictions on traveling or targeted
prophylaxis measures are implemented during the outbreak. 
We group the states in the nine influenza surveillance 
regions which are identical to the nine divisions of the US census and 
we use two different visualization strategies. In the first set of 
maps, regions are drawn with their normal size and a color code gives 
the prevalence of the infection in each region, i.e. the fraction of infected
individuals. This representation 
readily shows the high heterogeneity of the pandemic evolution. While 
useful such a visualization might be misleading, since the same 
prevalence obtained in different regions might correspond to very 
different values in the number of infected individuals if the two 
regions are very differently populated. Moreover, it is common to find 
strong population density heterogeneities, and it is not easy to 
detect visually a large level of contamination in a small but densely 
populated geographical area. In order to obtain a geographical 
representation which is able to carry at the same time information 
both on the level of infection and on the infection cases in each 
region, we have constructed the corresponding cartograms of the 
original maps in which the size of each geographic region (in our case 
US influenza surveillance regions) is rescaled according to its 
population. Several methods for constructing cartograms have been 
developed (see \cite{Gastner:2004} and references therein) and here we 
have adopted the diffusion-based method~\cite{Gastner:2004}, which 
produces cartograms by equalizing the population density through a 
linear diffusion process.  The geographical map representation 
readily shows the heterogeneity of the spatio-temporal epidemic
evolution but a quantitative characterization of this heterogeneity 
and its relation with the air transportation 
network statistical properties are major issues that are not yet 
fully explored.\\

\noindent  
{\bf Epidemic heterogeneity and network structure} 

In order to discriminate the role of the network structure on the 
spatio-temporal pattern of the epidemic process, 
we aim at a more quantitative analysis of its global heterogeneity.
This heterogeneity might find its origin just in the
stochastic nature of the infectious process or  determined by
the structural properties of the transportation network. In the latter
case, it is possible to envision the possibility of a larger
predictability of the epidemic behavior that would reflect the
underlying network structure. 
Here we introduce for the first time a characterization of the
epidemic pattern by using the entropy, a quantity customarily used in  
information theory to quantify the level of disorder of a signal or system.
At each time step, a snapshot of the epidemic pattern is provided 
by the set of values of the prevalence $i_j(t)=I_j(t)/N_j$ in each city $j$. 
We can therefore define the normalized vector $\vec{\rho}$ with 
components $\rho_j=i_j/\sum_\ell i_\ell$ which contains the
relevant information on the epidemic pattern. In particular, we can
measure the level of heterogeneity of the disease prevalence by
measuring the disorder encoded in the vector $\vec{\rho}$ with  
the normalized entropy function $H$ 
\begin{equation}  
H(t) = - \frac{1}{\log V} \sum_j \rho_j(t) \log  \rho_j(t) \ . 
\label{eq:entropy}  
\end{equation}  
If the epidemics is homogeneously affecting all nodes (i.e. all
prevalences are equal) the entropy attains its maximum value $H=1$.
Starting from $H=0$ which corresponds to one initial infected city -
the most localized and heterogeneous situation - $H(t)$ increases as
more cities become infected thus reducing the level of heterogeneity
(see Fig.~3{\bf A}). It is important to stress that in the present
context the entropy does not have any thermodynamical meaning. It must
be just considered as the appropriate mathematical tool able to
quantify the statistical disorder of a complicate spatio-temporal
signal.

\begin{figure}
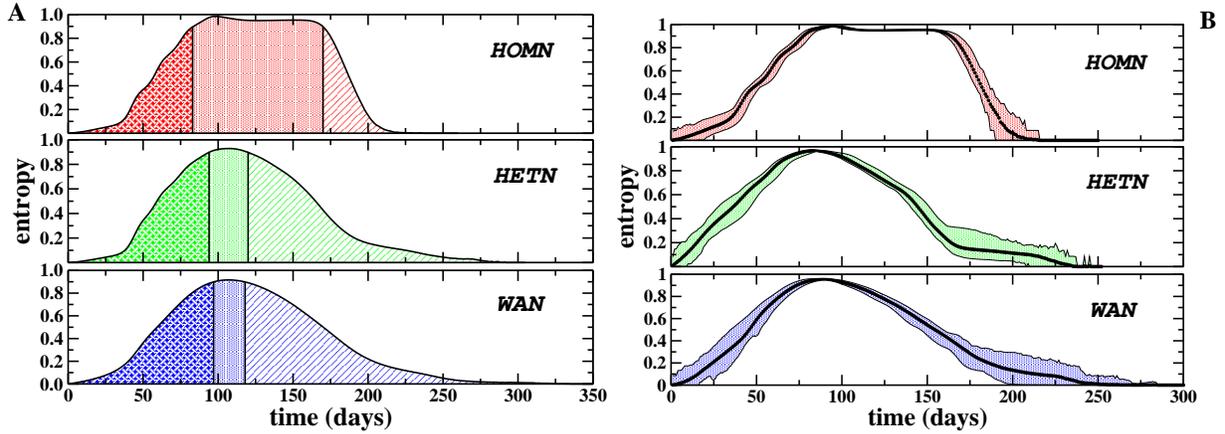
 

\centerline{ 
\includegraphics*[width=0.5\textwidth]{Hrho_m} 
\includegraphics*[width=0.5\textwidth]{Hrho} 
}

\caption{ 
Analysis of the heterogeneity of the epidemic pattern in the 
actual network (\emph{WAN}) compared with the two network 
models (\emph{HOMN}) and (\emph{HETN}). An SIR dynamics is adopted within
each city. 
{\bf (A)} Entropy $H(t)$ averaged over distinct initial 
infected cities and over noise realizations. Each profile is divided 
into three different phases, the central one corresponding to $H > 0.9$, i.e. 
to a homogeneous geographical spread of the 
disease. This phase is much longer for the \emph{HOMN} than 
for the real airport network. The behavior observed in 
\emph{HETN} is close to the real case meaning that the connectivity 
pattern plays a leading role in the epidemic behavior.  
{\bf (B)} Average value of the entropy, with the 
maximal dispersion obtained from $2\cdot 10^2$ noise realizations of 
an epidemics starting in Hong Kong. Fluctuations have a mild effect in 
all cases.} 
\label{fig:entropy} 
\end{figure}

To ascertain the effect of the network structure we compare the 
results obtained on the actual network with those obtained on 
different network models providing null hypotheses (see 
Fig.~\ref{fig:entropy}).  The first model network we consider (called 
\emph{HOMN}) is a homogeneous Erd\"os-R\'enyi random graph with the 
same number of vertices $V$ as the \emph{WAN}, and is obtained as 
follows: for each pair of vertices $(j,\ell)$, an edge is drawn 
independently, with uniform probability $p=\langle k \rangle /V$ where 
$\langle k \rangle$ is the average degree of the \emph{WAN}. In this 
way, we obtain a typical instance of a random graph with a poissonian 
degree distribution, peaked around the average value $\langle k 
\rangle$ and decreasing faster than exponentially at large degree 
values, in strong contrast with the true degree distribution of the 
\emph{WAN}.  For the second model (called \emph{HETN}) instead, we retain the 
exact topology of the real network. In both models, fluxes and 
populations are taken as uniform and equal to the corresponding 
averages in the actual air-transportation network. 
 
\begin{figure}

\centerline{ 
\includegraphics*[width=10cm]{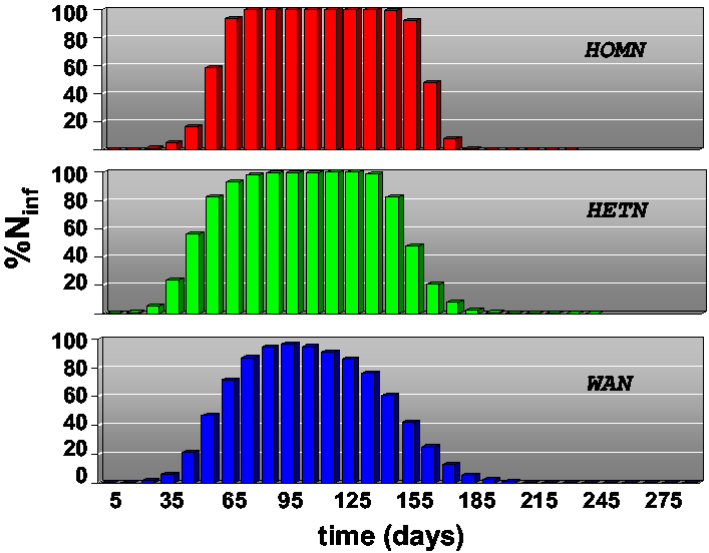} 
} 
\caption{ 
Percentage of infected 
cities as a function of time for an epidemics starting in Hong Kong
 based on an SIR dynamics within each city.
 The \emph{HOMN} case displays a large 
interval in which all cities are infected. The \emph{HETN} and 
the real case show a smoother profile with long tails, signature 
of a long lasting geographical heterogeneity of the epidemic 
diffusion. 
} 
\label{fig:Ninf} 
\end{figure}

The
differences in the behavior observed in the \emph{HOMN}, the
\emph{HETN} and in the real case provide striking evidence for a
direct relation between the network structure and the epidemic
pattern. The homogeneous network displays a homogeneous evolution
(with $H\approx 1$) of the epidemics during a long time window, with
sharp changes at the beginning and at the end of the spread. We
observe a different scenario for heterogeneous networks where $H$ is
significantly smaller than one most of the time, with long tails
signalling a long lasting heterogeneity of the epidemic behavior.
Indeed, the analytical inspection of the epidemic equations points out
that the broad variability of the contact pattern (degree
distribution) and the ratios $w_{j\ell}/N_j$ play an important role in the
heterogeneity of the spreading pattern.
 Strikingly, the curves obtained for both the real network
and the \emph{HETN} are similar, pointing out that in the case of the
airport network the broad nature of the degree distribution determines
to a large extent the overall properties of the epidemic pattern. 
Figure~\ref{fig:entropy}{\bf B} reports 
the average entropy profile together with the maximal dispersion
obtained for the spreading
starting from a given city  with different realizations of the noise. 
It is clear that the noise has a mild effect
and that the average behavior of the entropy is representative of
the behavior obtained in each realization.  In Fig.~\ref{fig:Ninf}
we show the percentage of infected cities as a function of time
for each null model and for the real case. While the \emph{HOMN}
displays a long time window in which all cities are infected, this
interval is much smaller in the \emph{HETN} and 
completely absent in the \emph{WAN}.\\

\noindent  
{\bf Predictability and forecast reliability}   

\begin{figure}[t] 
\centerline{ 
\includegraphics*[width=1.0\textwidth]{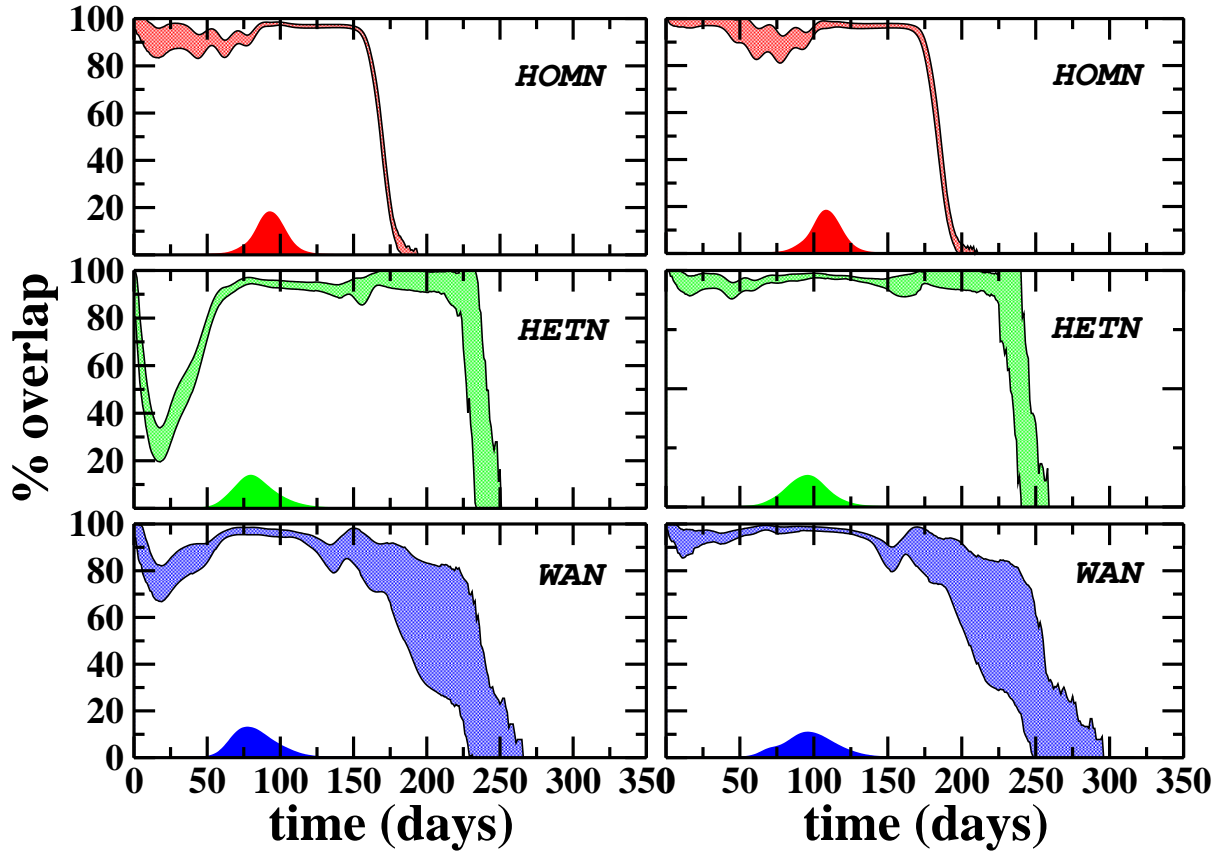} 
} 
\caption{ 
Percentage of overlap as a function of time: the shaded area corresponds to 
the standard deviation obtained with $5\cdot 10^3$ couples of different 
realizations of the global spreading model based on an SIR dynamics 
within each city. Topological heterogeneity plays a dominant role in reducing
the overlap in the early stage of the epidemics.
We observe two different behaviors depending on the
degree of the initially infected city: 
a reduced initial predictability in the case of airport hubs (left)
with respect to poorly connected cities (right). Large fluctuations at the end of the 
epidemics are observed  in 
the \emph{HETN} and in the real case, due to the different lifetime of the 
epidemics  in distinct realizations induced by the heterogeneity of 
the network.
We also report the prevalence profile as a function of time
showing that the maximum predictability corresponds to a prevalence
peak.} 
\label{fig:overlap} 
\end{figure}

A further major question in the modeling of global epidemics consists
in providing adequate information on the reliability of the obtained 
epidemic forecast; i.e. the epidemic predictability. Indeed the 
intrinsic stochasticity of the epidemic 
spreading will make each realization unique and reasonable forecast
can be obtained only if all epidemics outbreak realizations starting 
with the same initial conditions and subject to different noise 
realizations are reasonably similar. A convenient quantity to
monitor in this respect is the vector $\vec{\pi}(t)$ whose components are
$\pi_j(t)= I_j(t)/\sum_l I_l$; i.e the normalized probability that an
infected individual is in city $j$. The similarity between two
outbreaks realizations is quantitatively measured by the statistical
similarity of two realizations of the global epidemic characterized by
the vectors $\vec{\pi}^I$ and $\vec{\pi}^{II}$ respectively.  As a
measure of statistical similarity $sim(\vec{\pi}^I,\,\vec{\pi}^{II})$ we have considered the standard
Hellinger affinity $sim(\vec{\pi}^I,\,\vec{\pi}^{II})=\sum_j
\sqrt{\pi^I_j\pi^{II}_j}$. Normalized similarity measures do not
account for the difference in the total epidemic prevalence and we
have to consider also $sim(\vec{i}^I,\,\vec{i}^{II})$ where
$\vec{i}^{I(II)}=(i^{I(II)},1-i^{I(II)})$ and $i(t)=\sum_jI_j(t)/\cal{N}$ is
the worldwide epidemic prevalence ($\mathcal{N}=\sum_j N_j$ is the total
population). We can thus define the overlap function
measuring the similarity between two different outbreak realizations as
\begin{equation}  
\Theta(t)=sim\left(\vec{i}^I(t),\,\vec{i}^{II}(t)\right)\times sim\left(\vec{\pi}^I(t),\,\vec{\pi}^{II}(t)\right)
\label{eq:overlap}  
\end{equation}  
The overlap is maximal ($\Theta(t)=1$) when the
very same cities have the very same number of infectious individuals in both
realizations, and $\Theta(t)=0$ if the two realizations do not have
any common infected cities at time $t$. Clearly, a large overlap
corresponds to a predictable evolution, providing a direct measure of
the reliability of the epidemic forecast.  
In the \emph{HOMN} we find a significant overlap ($\Theta>80\%$, see
Fig.~\ref{fig:overlap}) even at the early 
stage of the epidemics - the most relevant phase for epidemic surveillance. 
The picture is different if we consider the \emph{HETN} and the 
real airport network where especially at the initial stage of the
epidemics the predictability is much smaller. 
These results may be rationalized by relating the level of predictability to
the presence of a backbone of dominant spreading channels defining
specific ``epidemic pathways'' which are weakly affected by the
stochastic noise. Epidemic pathways are the outcome of the conflict 
between two different properties of the network. On the one hand, the  heterogeneity
of the connectivity pattern provides a multiplicity of equivalent
channels for the travel of infected individuals depressing the
predictability of the evolution. On the other hand, the heterogeneity
of traffic flows introduces dominant connections which select
preferential pathways increasing the epidemic predictability. 
The heterogeneous connectivity pattern of the \emph{HETN} and the \emph{WAN} thus
generates a multiplicity of channels that decreases the predictability. 
In the real case the  lowering of the epidemic predictability
also indicates the dominant effect of  the topological
heterogeneity that wins over the opposite tendency of the traffic \
heterogeneity. The above framework is confirmed by the two distinct behaviors  
depending on the degree of the initial infected city. Epidemics
starting in  initial cities
with a hub airport generate realizations whose overlap initially
decreases to 50-60$\%$ because of the many possible equivalent paths
resulting  in a larger
differentiation of the epidemic history in each stochastic
realization. On the contrary, outbreaks from poorly connected initial
cities display a large overlap due to the few available connections
that favor the selection of specific epidemic pathways.~\\ 
  
\noindent  
{\bf Outlook}  
   
From our study, it emerges that the air transportation network  
properties are responsible of the global pattern of emerging  
diseases. In this perspective, the complex features characterizing this  
network are the origin of the heterogeneous and seemingly erratic  
spreading on the global scale of diseases such as SARS.  The analysis  
provided here show that large scale mathematical models that takes  
fully into account the complexity of the transportation matrix can be  
used to obtain detailed forecast of emergent disease outbreaks. 
We have also shown that it is possible to provide quantitative 
measurements of the predictability of epidemic patterns, providing a 
tool that might be used to obtain confidence intervals in epidemic  
forecast and in the risk analysis of containement scenarios. 
It is clear that to make the forecast more realistic, it is necessary to  
introduce more details in the disease dynamics. In particular, seasonal  
effects and geographical heterogeneity in the basic transmission rate  
(due to different hygienic conditions and health care systems in  
different countries) should be addressed. Finally, the interrelation of  
the air transportation network with other transportation systems such  
as railways and highways could be very useful for forecast on longer  
time scales. We believe however that the basic understanding of the  
interplay of the transportation network complex features with the  
disease spreading evolution and the detailed modeling obtained by the  
full consideration of these features may represent a valuable tool to  
test traveling restrictions and vaccination policies in the case of  
new pandemic events.~\\  
  
\noindent   
{\small We thank IATA for making the airline commercial flight database  
  available to us. M.B. is on leave of absence from 
CEA, 
D\'epartement de Physique Th\'eorique et 
Appliqu\'ee BP12, 91680 Bruy\`eres-Le-Ch\^atel, France.}

%

\vspace{-.5cm}

\begin{thebibliography}{99}  
 
\bibitem{May:1992} Anderson, R.M. \& May, R.M. (1992), {\it Infectious 
diseases in humans} (Oxford University Press, Oxford). 
 
\bibitem{Hethcote:1984} Hethcote, H.W. \& Yorke, J.A. (1984), 
{\it Lect. Notes Biomath.} {\bf 56} (Berlin, Springer-Verlag). 
 
\bibitem{Morris:1996} Kretzschmar, M \& Morris, M. (1996), 
{\it Math. Biosci.}, {\bf 133}, 165-195. 
 
\bibitem{Keeling:1999} Keeling, M. (1999), 
{\it Proc. R. Soc. Lond. B} {\bf 266}, 859-867. 
 
\bibitem{Pastor:2001} Pastor-Satorras, R. \& Vespignani, A. (2001), 
{\it Phys. Rev. Lett.} {\bf 86}, 3200-3203. 
 
\bibitem{May:2001} Lloyd A.L. \& May, R.M. (2001), 
{\it Science}  {\bf 292}, 1316-1317. 

\bibitem{Meyers:2005} Meyers, L.A , Pourbohloul B., Newman, M.E.J., 
Skowronski, D.M. 
\& Brunham,R.C. (2005), Journal of Theor. Biol. {\bf 232} 71-81.
 
\bibitem{Ferguson:2003} Ferguson, N.M. {\it et al} (2003),  
{\it Nature} {\bf 425}, 681-685.  
 
\bibitem{Chowell:2003} Chowell, G., Hyman, J.M., Eubank, S. \& 
Castillo-Chavez, C. (2003), 
{\it Phys. Rev. E} {\bf 68}, 066102. 
 
\bibitem{Eubank:2004} Eubank, S., Guclu, H., Anil Kumar, V.S., 
Marathe, M.V., Srinivasan, A., Toroczkai, Z. \& Wang, N. (2004), 
{\it Nature} {\bf 429}, 180-184. 
 
 
\bibitem{Barabasi:2000} Albert, R. \& Barab{\'a}si, A.-L. (2000), 
{\it Rev. Mod. Phys.} {\bf 74}, 47--97. 
 
 
\bibitem{Doro:2003} Dorogovtsev, S.N. \& Mendes, J.F.F. (2003),  
{\em Evolution of networks: From 
biological nets to the {I}nternet and {WWW}} (Oxford University Press, 
Oxford). 
 
\bibitem{Pastorbook:2003} Pastor-Satorras, R. \& Vespignani, A. (2003), 
{\em Evolution and structure of the Internet: A statistical physics 
approach} (Cambridge University Press, Cambridge). 
  
\bibitem{Cohen:2000} Cohen, M.L. (2000),  
{\it Nature} {\bf 406}, 762-767.  
 
\bibitem{Cliff:2004} Cliff, A. \& Haggett, P. (2004), 
{\it British Medical Bulletin}, {\bf 69}. 87-99.   
   
\bibitem{SARS} http://www.who.int/csr/sars/en 
 
\bibitem{Longini:1985}  
Rvachev, L.A. \& Longini, I.M. (1985), 
{\it Mathematical Biosciences} {\bf 75}, 3-22.  
 
\bibitem{Baroyan:1969} 
Baroyan, O.V., Genchikov, L.A., Rvachev, L.A, Shashkov, V.A. (1969), 
{\it Bull. Internat. Epidemiol. Assoc.}, {\bf 18}, 22-31. 
 
\bibitem{Longini:1988} Longini, I.M. (1988), 
{\it Mathematical Biosciences} {\bf 90}, 367-383. 
 
\bibitem{Grais:2003} Grais, R.F., Hugh Ellis, J., \& Glass, G.E. (2003), 
{\it European Journal of Epidemiology} {\bf 18}, 
1065-1072. 
 
\bibitem{Grais:2004} Grais, R.F., Hugh Ellis, J., Kress, A. \& Glass, G.E. 
(2004), 
{\it Health Care Management Science}, {\bf 7}, 127-134. 
 
\bibitem{Flahault:1991} Flahault, A. \& Valleron, A.-J. (1991), 
{\it Math. Pop. Studies} {\bf 3}, 1-11. 
 
\bibitem{Hufnagel:2004} Hufnagel, L., Brockmann, D. \& Geisel, T. (2004),  
{\it Proc. Natl. Acad. Sci. }(USA) {\bf 101}, 15124-15129.  
     
\bibitem{Barrat:2004} Barrat, A., Barth{\'e}lemy, M., Pastor-Satorras, R. 
\& Vespignani, A. (2004), 
{\it Proc. Natl. Acad. Sci.} USA {\bf 101}, 3747-3752.  
 

\bibitem{Amaral:2004} Guimer\`a, R., Mossa, S.,  Turtschi, A.,  Amaral, L.A.N.,
 {\it Proc. Natl. Acad. Sci. USA}  {\bf 102}, 7794 (2005).
 
\bibitem{IATA} http://www.iata.org  
 
\bibitem{Zipf:1949} Zipf, G.K., {\it Human behavior and the principle   
of least efforts} (Addison-Wesley, 1949).  
 
\bibitem{Gardiner} Gardiner W.C., {\it Handbook of stochastic methods 
for physics, chemistry and natural sciences}, Springer, 3rd edition, 
2004. 
 
\bibitem{Gillespie} Gillespie, D.T.  (2000), 
{\it J. Chem. Phys.}, {\bf 113}, 297-306. 

\bibitem{MarroDickman}  
Marro, J. \& Dickman, R. (1998), {\it Nonequilibrium phase transitions  
and critical phenomena}, Cambridge University Press, Cambridge. 
  
\bibitem{Dickman:1994}  
Dickman, R. (1994), 
{\it Phys. Rev. E} {\bf 50}, 4404-4409.  

\bibitem{Gastner:2004} Gastner, M.T. \& Newman, M.E.J. (2004), 
{\it Proc. Natl. Acad. Sci. (USA)} {\bf 101}, 7499-7504.  
  


 
 
  
 
 
  
\end{thebibliography}
\end{document}